\documentclass[sigconf]{acmart}
\usepackage{hyperref}
\usepackage{url}
\usepackage[tight,footnotesize]{subfigure}
\usepackage{balance}
\usepackage{multirow}
\usepackage{fancybox}
\usepackage{mathrsfs}
\usepackage{wrapfig}
\usepackage{graphicx}
\usepackage{amsmath}
\usepackage{mathtools}
\usepackage[mathscr]{euscript}
\usepackage{algorithm}
\usepackage{algorithmic}

\usepackage{amsthm}

\theoremstyle{definition}
\theoremstyle{remark}

\graphicspath{{figures/}}
\usepackage{ctable}
\usepackage{bbm}
\usepackage{verbatim}
\usepackage{comment}

\usepackage{booktabs} 

\usepackage{fancyhdr,lipsum}
\setcopyright{rightsretained}

\begin{document}

\thispagestyle{plain}
\title{Bayesian Non-Exhaustive Classification for Active Online Name Disambiguation}

\author{Baichuan Zhang}
\affiliation{%
  \institution{Purdue University}
}
\email{zhan1910@purdue.edu}

\author{Murat Dundar}
\affiliation{%
  \institution{Indiana University Purdue University Indianapolis}
}
\email{mdundar@iupui.edu}

\author{Mohammad Al Hasan}
\affiliation{%
  \institution{Indiana University Purdue University Indianapolis}
}
\email{alhasan@cs.iupui.edu}

\begin{abstract}

The name disambiguation task partitions a collection of records pertaining to
a given name, such that there is a one-to-one correspondence between the 
partitions and a group of people, all sharing that given name. Most existing 
solutions for this task are proposed for static data. However, more realistic scenarios stipulate emergence of records in a streaming fashion where records may belong to known as well as unknown persons all sharing the same name. This requires a flexible name disambiguation algorithm that can not only classify records of known persons represented in the training data by their existing records but can also identify records of new ambiguous persons with no existing records included in the initial training dataset. Toward achieving this objective, in this paper we propose a Bayesian non-exhaustive classification framework for solving online name disambiguation. In particular, we present a Dirichlet Process Gaussian Mixture Model (DPGMM) as a core engine for online name disambiguation task. Meanwhile, two online inference algorithms, namely one-pass Gibbs sampler and Sequential Importance Sampling with Resampling (also known as particle filtering), are proposed to simultaneously perform online classification and new class discovery. 
As a case study we consider bibliographic data in a temporal stream format and disambiguate authors by partitioning their papers into homogeneous groups.Our experimental results demonstrate that the proposed method is significantly better than existing methods for performing online name disambiguation task. We also propose an interactive version of our online name disambiguation method designed to leverage user feedback to improve prediction accuracy.  

\end{abstract}

\maketitle


\section{Introduction}~\label{sec:intro}
%
%
Popular names are shared by many people around the world. When such names
are mentioned in various on-line and off-line documents, more often, ambiguity arises; i.e., 
we cannot easily deduce from the document context which real-life person a given mention is being referred to. Being unable to resolve this ambiguity often leads to erroneous aggregation of documents of multiple persons who are namesake of one another. Such mistakes deteriorate the performance of document retrieval, web search, and bibliographic data analysis. For bibliometrics and library sciences, many  distinct authors in the academic world share the same name. As a result, the bibliographic servers that maintain publication data may mistakenly aggregate the articles from multiple scholars (sharing the same name) into a
unique profile in some digital repositories. For an example, the Google scholar profile
associated with the name ``Yang Chen" (GS)~\footnote{\url{https://scholar.google.com/citations?user=gl26ACAAAAAJ&hl=en}}
is verified as the profile page of a Computer Graphics PhD candidate at Purdue, but based 
on our labeling, more than
20 distinct persons' publications are mixed under that profile mistakenly. Such issues in library science over- or under-estimate a researcher's citation related impact metrics.
Beyond the academic world, name ambiguity conundrum also causes misidentification during
counter-terrorism efforts, leading to severe distress to many individuals  who happen to share names with  wanted suspects. 

{\em Name disambiguation} task is used to resolve the name ambiguity problem. Formally speaking, given a large collection of records pertaining to a single name value, the name disambiguation task
partitions the records into groups of records, such that each group belongs to a unique real-life person. In the above definition
the term ``record'' refers to any form of collective information associated with the mention of a given name. For instance, in a digital repository of academic publications, a record is simply the citation context (title, co-authors, and venue) of a paper.  In case of a mention of a name in an online news article, a record may include information such as, article title, sentence context of the mention, and other associated name references within the article. For a 
social network profile, the record may contain publicly available friend-list,
and text from the posts in that profile.
A record does not typically include structured biographical information, 
such as, address, DOB, SSN, which can disambiguate a person  instantly by using
an `and' query.

Due to its wide-spread applications, and difficulty, building machine learning based methods for resolving
name ambiguity has attracted continuous attention from researchers in information 
retrieval, natural language processing, and data mining
communities~\cite{Han.Giles.ea:04,Tang.Fong.ea:12,Cen.Luo.ea:13,Li2012,Hoffart:2011:RDN:2145432.2145521,Han.Zhao:09,Zwicklbauer.Seifert.ea:16,emnlp.jietang}. The
principle approach of these existing methods is clustering or supervised
classification, for which the number of classes (or clusters) is known beforehand.
Besides, the
majority of the existing approaches to name disambiguation operate in a batch mode, where all the records to be disambiguated are assumed to be accessible to the algorithm initially. This assumption requires to run a new  disambiguation task every time a record is added to the collection. However, due to the fast growth of digital library, or streaming data sources (Twitter, Facebook), rerunning disambiguation process on the whole data every time a new record is added would not be very economical. Instead, it is more practical to perform this task in an incremental fashion by considering the streaming nature of records. 

\begin{figure}
\vspace{-0.40in}
\includegraphics[height=0.70\linewidth] {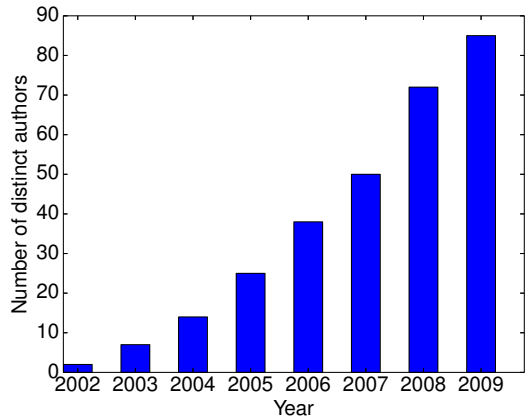}
\caption{Name Ambiguity Evolution for the name ``Jing Zhang''}
\vspace{-0.20in}
\label{fig:intro1}
\end{figure}

Designing an incremental, i.e., online, name disambiguation is challenging as the method must be able to adapt to a {\em non-exhaustive} training dataset~\footnote{
A training dataset is called exhaustive if it contains records 
for all values (classes) of the target variable, otherwise it is called non-exhaustive.
}. In other words, it should be able to identify records belonging to new ambiguous persons who do not have any historical records in the system.  After identification,
the learning algorithm must re-configure the model (for instance, update the number of classes, $k$) so that it can  correctly recover future records of this newly found ambiguous person. This is an important requirement because in real-life, for a common name, a significant number of streaming records belongs to novel (not yet seen) persons sharing that name.  As an example, consider the name reference ``Jing Zhang" from Arnetminer~\footnote{\url{http://arnetminer.org}},
a well known bibliographic dataset. As shown in figure~\ref{fig:intro1}, 
the number of distinct real-life authors in Arnetminer sharing the name ``Jing Zhang"
has increased from $14$ to $85$ between the year $2004$ and $2009$.
Evidently, the training dataset of online name
disambiguation is never exhaustive and any supervised classifier
trained on the assumption of exhaustive training dataset mis-classifies (with certainty) 
all the records belonging to a novel ambiguous person. 

Besides non-exhaustiveness, {\em online verification} is another desirable property for an incremental name
disambiguation system.  Such a system asks users to provide feedback on the correctness
of its prediction. Feedback collection can be automated by using online social networks 
or crowdsourcing platforms.  As an example, consider the online digital library platform ResearchGate~\footnote{\url{https://www.researchgate.net/}}; it performs author name disambiguation
by asking a potential researcher whether he is the author of a paper before adding that paper to that person's profile. Human feedback significantly improves the accuracy of  a name disambiguation task;  
however, to reduce human effort the system should consult the human as infrequently as possible, and the consultation should be made for documents, for which the human feedback would yield the maximum utility for reconfiguring the model. Thus, designing an active name
disambiguation system that can accommodate streaming non-exhaustive data is another focus of this work.

A few works that perform online name disambiguation by considering
non-exhaustive training dataset have emerged recently. For example, Khabsa et al.~\cite{Khabsa.Giles.ea:15} propose a DBSCAN based density estimation model to classify new records while they are being added incrementally. Qian et al.~\cite{Qian.Zheng.ea:15} present a probabilistic model to determine the class membership of a newly added record. Ariano et al.~\cite{Adriano.Anderson.ea:12} introduce an association rule based approach for detecting unseen authors. Even though all these studies are able to adapt to the non-exhaustive scenario, their corresponding online prediction models use heuristically
chosen threshold values to decide whether a record belongs to a new ambiguous author or not; such approaches are highly susceptible to the choice of threshold parameters. 


In this work we propose a new Bayesian non-exhaustive classification framework for active online name disambiguation problem. Our method uses a Dirichlet Process Gaussian Mixture Model (DPGMM) as the core engine. The DPGMM facilitates online non-exhaustive
classification by being partially-observed,
where existing classes are modeled by observed components and emerging/future classes by the unobserved ones. The hyperparameters of the base distribution of Dirichlet process, which is chosen as a bivariate Normal $\times$ Inverted Wishart (NIW), enables information sharing across existing as well as emerging classes. The hyperparameters 
are estimated using offline records initially accessible in the training set. 
For prediction (i.e., assigning class label to a test record), we propose two independent online inference mechanisms, the first is based on  one-pass Gibbs sampler
and the second is based on particle filtering. For both cases, the inference process jointly tackles online classification and emerging class discovery i.e., the inference
process evaluates the probability of assigning a future record to an emerging class or to one of the existing ones. It also uses predicted label information to update the hyperparameters of our online model and thus adapt the new classification model to classify subsequent records.
We also extend our method to active online name disambiguation  task where the method systematically selects a small
number of records and seeks user feedback regarding their true class of origin in an effort to effectively reconfigure the model. 

%
%
An earlier version of this paper has been published
as a conference article~\cite{zhang2016bayesian}. However, the earlier version only included Gibbs sampling as the inference
mechanism. In this version we have additionally proposed 
particle filtering based inference, which is much improved
over the Gibbs sampling based inference, as shown in the
experimental result section. Also, this version incorporates
active learning within the inference process, which was absent
in the conference version of this paper.

Below we summarize the contributions of this work:

\begin{enumerate}

\item We study online name disambiguation problem in a non-exhaustive streaming setting and propose a self-adjusting Bayesian non-exhaustive model that is capable of performing online classification, and novel class discovery at the same time. To the best of our knowledge, our work is the first one to adapt Bayesian non-exhaustive classification for online name disambiguation task.

\item We propose two online inference algorithms, namely one-pass Gibbs sampler and particle filtering, for Dirichlet Process Gaussian Mixture Model to perform online non-exhaustive classification in order to efficiently evaluate the class assignment of an online record.  

\item We enhance our proposed online name disambiguation approach by making it interactive, so that user guidance can be incorporated to improve the disambiguation performance. To the best of our knowledge, our model is the first work where active learning is coupled with non-exhaustive online learning for name disambiguation.
%
%
%
%
%

\item We use bibliographic datasets to evaluate the proposed approach against several benchmarks. The results demonstrate the superiority of the proposed approach over the state-of-the-art  in online name disambiguation.

\end{enumerate}

\section{Related Work}~\label{sec:rw}

There is a large body of work on name disambiguation in the literature. In terms of methodologies,
supervised~\cite{Bunescu.Pasca:06, Han.Giles.ea:04}, unsupervised~\cite{Han.Zha.ea:05,Cen.Luo.ea:13}, and probabilistic
relational models~\cite{Tang.Fong.ea:12, Wang.Tang.ea:11,Song.Huang.ea:07}, are considered. In a supervised setting, a distinct
entity can be considered as a class, and the objective is to classify each record to one of the classes.
Han et al.~\cite{Han.Giles.ea:04} propose supervised name disambiguation methodologies by utilizing Naive Bayes and 
SVM. For
unsupervised name disambiguation, the records are partitioned into several clusters with the goal of
obtaining a partition where each cluster contains records from a unique entity. For example, ~\cite{Han.Zha.ea:05}
proposes one of the earliest unsupervised name disambiguation methods for bibliographical data, which is based on $K$-way spectral
clustering. Specifically, the authors compute Gram matrix representing similarities between different citations and 
apply $K$-way spectral clustering algorithm on the Gram matrix in order to obtain the desired
clusters of the citations. Recently, probabilistic relational models, especially graphical models have also been
considered for the name disambiguation task. For instance,~\cite{Tang.Fong.ea:12} proposes to use Markov
Random Fields to address name disambiguation challenge in a unified probabilistic framework. Another work
is presented in ~\cite{Wang.Tang.ea:11} which uses pairwise factor graph model for this task.
~\cite{Zhang.Saha.ea:14, Zhang.Hasan.ea:15, Hermansson.Kerola.ea:13,Zhang.Hasan.3} present approaches for the name disambiguation task on
anonymized graphs and they only leverage graph topological features due to the privacy concern. 
%
%
In addition, ~\cite{zhang2017feature} formalizes name disambiguation problem as a privacy-preserving
classification task such that the anonymized dataset satisfies non-disclosure requirements while achieving high disambiguation performance.
A survey article is also available, which presents a taxonomy of various name disambiguation
methods in the existing literature~\cite{Ferreira.Laender.ea:12}. 

Most of existing methods above tackle disambiguation task in a batch setting, where all records to be resolved
are initially available to the algorithm, which makes these techniques unsuitable for disambiguating a future record. In
recent years, online name disambiguation was considered in a few works ~\cite{Qian.Zheng.ea:15,Khabsa.Giles.ea:15,Adriano.Anderson.ea:12,Ana.Alberto.ea:11}. These techniques 
perform name disambiguation incrementally without the need to retrain the system every time a new record is received.
Khabsa et al.~\cite{Khabsa.Giles.ea:15} use an online variant of DBSCAN, a well-known density-based clustering
technique to cluster new records incrementally as they become available. Since, DBSCAN does not use a fixed number of clusters it can adapt to the non-exhaustive scenario by simply assigning a new record to a new cluster, as needed. However, DBSCAN is quite susceptible to the choice of parameter values, and depending on the specific values chosen, a record of an emerging class can be simply labeled as an outlier instance.
~\cite{Qian.Zheng.ea:15} proposes a two stage framework for online name disambiguation. The first
stage performs batch name disambiguation to disambiguate all the records that appeared no later than a given time threshold using hierarchical agglomerative clustering. The second stage performs
incremental name disambiguation to determine the class membership of a newly added record. However, the method uses a
heuristic threshold to decide on the cluster assignments of new records which makes the performance of this approach very sensitive  to the choice of threshold parameter.
~\cite{Adriano.Anderson.ea:12} introduces an association rule
based approach for detecting unseen authors in test set.  The major drawback of their proposed solution is
that it can only identify records of emerging authors in a binary setting but fails to further distinguish among
them. Besides, the approach is not very robust with respect to the threshold parameter used in the association rule discovery. 

Another line of work approaches name disambiguation from an active learning perspective~\cite{Huang.Lee.ea:06,Wang.Tang.ea:11,Cheng:2013:BAN:2505515.2507858}. For example, authors in~\cite{Huang.Lee.ea:06} propose a method that queries the label information for the most ambiguous records. Authors in ~\cite{Wang.Tang.ea:11} present a pairwise factor graph model for active name disambiguation, which maximizes the utility of user's corrections for improving the disambiguation performance. Another recent work uses crowdsourcing for active name disambiguation~\cite{Cheng:2013:BAN:2505515.2507858}. However, all of these active name disambiguation techniques proposed in the offline setting. 

Our proposed solution utilizes non-exhaustive learning---a rather recent development in machine learning
~\cite{Ferit.Dundar.ea:10, Dundar.Akova.ea:12, Miller.Browning:03}. Akova et al.~\cite{Ferit.Dundar.ea:10} propose a Bayesian  approach for detecting emerging classes based on posterior probabilities. However, the decision function 
for identifying emerging classes uses a heuristic threshold and does not consider a prior model over class 
parameters; hence the emerging class detection procedure of this model is purely data-driven. 
Miller et al. ~\cite{Miller.Browning:03} present a mixture model using expectation maximization (EM) for online class discovery. 
~\cite{Dundar.Akova.ea:12} is another non-exhaustive learning work for emerging class discovery and the work is mainly motivated by a bio-detection application. 

%
%
\section{Online Name Disambiguation Challenges}~\label{sec:challenge}

For a given name reference $a$, assume $R_{n}$ is a stream of records associated with $a$. The subscript
$n$ represents the identifier of the last record in the stream and the value of this identifier increases
as new records are observed in the stream. Each record $r_{i} \in R_{n}$
can be represented by a $d$-dimensional vector which is the feature representation of the record in a 
metric space. In real-life, the name reference $a$ is associated with multiple persons 
(say $k$) all sharing the same name, $a$. The task of name disambiguation is to partition 
$R_{n}$ into $k$ disjoint sets such that each partition contains records of a unique person entity.
When $k$ is fixed and known a priori, name disambiguation can be solved as a $k$-class classification
task using supervised learning methodologies. However, for many domains the number of classes ($k$)
is not known, rather with new records being inserted in the stream $R_{n}$ , the number of distinct 
person entities associated with $a$ may increase. The objective of online name disambiguation is to 
learn a model that assigns each incoming record into an appropriate partition containing records 
of a unique person entity.

Online name disambiguation is marred by several challenges, which we discuss below: 
%
%

First, for a given record stream $R_{n} = \{r_1, \cdots,  r_{i}, \cdots, r_n\}$, the record $r_i$ 
is classified with the records leading up to $r_{i-1}$, i.e. $R_{i-1}$ is our training data for
this classification task. However, the record $r_i$ may belong to a new person entity (having name $a$) 
with no previous records in $R_{i-1}$. This happens because for online
setting, the number of real-life name entities in $R_{n}$ is not fixed, rather it increases 
over the time. A traditional $k$-class supervised classification model which is trained with records 
of known entities mis-classifies the new emerging record with certainty, leading to an ill-defined classification 
problem. So, for online name disambiguation, a learning model is needed which works in non-exhaustive
setting, where instances of some classes are not at all available in the training data. In existing works,
this challenge is resolved using clustering framework where a new cluster is introduced for the emerging 
record of a new person entity,
but this solution is not robust because small changes in clustering parameters make widely varying
clustering outcomes.

The second challenge is that online name disambiguation, more often, leads to a severely imbalanced 
classification task. This is due to the fact that in most of the real-life name disambiguation problems, 
the size of the true partitions 
of the record set $R_{n}$ follows a power-law distribution. In other words, there are a few persons (dominant 
entities) with the
name reference $a$ to whom the majority of the records belong. Only a few records (typically one or two) belong
to each of the remaining entities (with name reference $a$). Typically, the persons whose records appear at earlier time
are dominant entities, which makes identifying novel entity an even more challenging task.

The third challenge in online name disambiguation is related to online learning scenario, where the incoming
record is not merely a test instance of typical supervised learning. Rather, the learning algorithm requires to detect
whether the incoming record belongs to a novel entity, and if so, the algorithm must adapt itself and configure model 
to identify future records of this novel entity. Overall, this requires a 
self-adjusting model that updates the number of classes to accurately classify incoming 
records to both new and existing classes. 

The final challenge in our list is related to temporal ordering of the records. In traditional classification,
records do not have any temporal connotation, so an arbitrary train/test split is permitted. But, for online
setting the model must respect time order of the records, i.e., a future record cannot be used for building a
training model that classifies older records.

Our proposed model overcomes all the above challenges by using a principled approach.

\section{Online Name Disambiguation on Bibliographic Data}

As we have mentioned earlier, name disambiguation is a severe issue in digital library domain.  In
many other domains, solving name disambiguation is easier as the method may have access to personalized 
attributes of an entity, 
such as institution affiliation, and email address. But, in digital library, the reference of a paper only includes paper
title, author name, publication venue, and year of publication, which are not sufficient for disambiguation of most of
the name references. Besides, in many citations the first name of the authors are often replaced by initials, which worsen 
the disambiguation issue. As a result, nearly, all the
online bibliographic repositories, including DBLP, Google scholar, ArnetMiner, and PubMed, suffer from this issue.
Nevertheless, these repositories provide timely update of the publication data along with their chronological orders, so they
provide an ideal  setting for evaluating the effectiveness of an online name disambiguation method.

In this work, we use bibliographic data as a case study for online name disambiguation. For each name reference $a$,
we build a distinct classification model. The record stream $R_{n}$ for the name reference $a$ is the chronologically ordered stream of scholarly publications where $a$ is one of the authors. To build a feature vector for a paper in
$R_{n}$ we extract features from its author-list, keywords from its paper title, and paper venue (journal/conference). We provide more details of feature construction in the following subsection.

\subsection{Feature Matrix Construction and Preprocessing}~\label{sec:preprocessing}

For a given name reference $a$, say we have a record stream containing $n$ papers for which the name reference 
$a$ is in the author-list. We represent each paper with a $d$ dimensional feature vector. Then we define a data matrix for $a$, in which each row represents a record and each column corresponds to a feature. In addition, each record has a class label $l_{i}$ to represent the $i$-th distinct person sharing the name reference $a$. The name disambiguation task is to partition $R_{n}$ into $k$ disjoint sets such that each partition contains records of a unique person with name reference $a$.  Note that, the $k$ value is not fixed, rather it can increase as emerging records appear in the stream $R_n$. 

Following earlier works on name disambiguation in bibliographic domain~\cite{Han.Giles.ea:04,Wang.Tang.ea:11,Tang.Fong.ea:12}, we use
coauthor information, publication and venue titles as features for a publication $r_{i}$. 
For coauthor information, we first partition the coauthor list of each paper 
(except $a$) into authors, 
then define a binary feature for each author (indicating existence or not). Paper titles are processed using standard NLP tools to remove all numbers, special characters as well as stop words. After that, we generate binary value for each of the remaining words in the title as its feature value.  Paper venue (name of journal or conference) is also a binary feature,
getting a value of 0 or 1 depending on whether the paper is published in a venue or not. 

To address the sparsity of the generated binary feature representation, we pre-process the data by using an incremental version of non-negative matrix factorization (INNMF), which sequentially embeds the original feature matrix into a low dimensional space denoted as $X_{n} \in {\rm I\!R}^{n \times h}$, where $h$ is the latent dimension. Specifically, we first perform NNMF~\cite{Daniel.Seung:01} in the batch mode using the initially available training records. Then for each online record, we represent it as a linear combination of a set of basis vectors generated from the training set. The coefficients serve as latent features for each online record. In order to learn the coefficients, we solve a constrained quadratic programming problem by minimizing
a least square loss function under the constraint that each coefficient is non-negative.
The justification of using INNMF is to discover effective latent feature representation for each online record to better fit our proposed Normal $\times$ Normal $\times$ Invert Wishart (NNIW) data model. Considering the above feature representation of the records, in subsequent discussion we will use $X_n$ to represent the records in $R_n$.

\subsection{Problem Formulation}

The active online name disambiguation problem is formally defined as follows: 
given a temporal partition $t_{0}$, we consider two types of records. First
type consists of a collection of $n$ records $X_{n} = \{ x_{i}\}_{i = 1}^{n}$ whose time-stamp is smaller
or equal to $t_0$. They serve as the training set 
 with known labels denoted as $Y_{n} = \{y_{i}\}_{i = 1}^{n}$,  where $y_{i} \in \{l_{1}, ..., l_{k}\}$, and $k$ is the number of distinct classes in the training set. The second type of records has time-stamp higher than $t_0$; they are represented as $\tilde{X}_{n^{u}} = \{\tilde{x}_{i}\}_{i = 1}^{n^{u}}$, $n^u$ ($u$ stands for unobserved) is the number of records sequentially observed online. The online
 name disambiguation task is to predict the labels of these records
denoted as $\tilde{Y}_{n^{u}} = \{\tilde{y}_{i}\}_{i = 1}^{n^{u}}$, where $\tilde{y}_{i} \in \{l_{1}, ..., l_{k + \tilde{k}_{n^{u}}}\}$ and $\tilde{k}_{n^{u}}$ is the number of emerging classes associated with $n^{u}$ online records. 

Given an online record $\tilde{x}_{i}$, our proposed model computes its probability for belonging to one of the existing classes or an emerging one. Based on the computed probability, if the disambiguation result is uncertain, we request the ground-truth label information of this particular online record from user and then re-configure the model for classifying subsequent records. Note that, user interactiveness is an added feature independent of prediction method. If the user feedback is unavailable,
the method simply predicts the label of a record based on its computed probability and proceeds thereon.

\section{Methodology}

In this section we discuss our proposed Bayesian non-exhaustive online name disambiguation methodologies. The methodologies discussed  in this section are domain neutral and can be applied to any domain, once an appropriately constructed feature matrix is obtained.

\subsection{Dirichlet Process Gaussian Mixture Model}~\label{sec:DPGMM}

The Dirichlet Process (DP)~\cite{Teh:10} is one of the most widely used Bayesian non-parametric priors, parameterized by a concentration parameter $\alpha > 0$ and a base distribution $H$ over a given space $\theta \in \Theta$. Although the base distribution $H$ can be continuous, a sample $G \sim DP(\alpha, H)$ drawn from a DP is a discrete distribution. In order to represent samples $G$ drawn from a DP, it is a common practice to use stick breaking construction~\cite{Sethuraman:94} as below: 

\begin{eqnarray}
\label{eq:sbc}
\phi_{i} &\sim& H \nonumber \\
\beta_{i} &\sim& Beta(1, \alpha) \nonumber \\
\pi_{i} &=& \beta_{i}\prod_{j=1}^{i-1}(1-\beta_{j}) \nonumber \\
\end{eqnarray}

As shown in Equation~\ref{eq:sbc}, in order to simulate the process of stick breaking construction, imagine we have a stick of length $1$ to represent total probability. 
We first generate each point $\phi_{i}$ from base distribution $H$, which originates from our proposed Normal $\times$ Invert Wishart data model.  
Then we sample a random variable $\beta_{i}$ from Beta(1, $\alpha$) distribution.  
After that we break off a fraction $\beta_{i}$ of the remaining stick as the weight of parameter $\phi_{i}$, denoted as $\pi_{i}$. 
In this way it allows us to represent random discrete probability measure $G$ as a probability mass function in terms of infinitely many 
$\phi_{1}, ..., \phi_{\infty}$ and their corresponding weights  $\pi_{1}, ..., \pi_{\infty}$ 
yielding $G = \displaystyle \sum_{i=1}^{\infty}\pi_{i}\delta_{{\phi_{i}}}$, where $\delta_{{\phi_{i}}}$ is the point mass of $\phi_{i}$.

Thanks to the discrete nature of $G$, DP offers an online clustering/classification of the streaming records as a by-product. Specifically, given a set of $n$ records $X_{n} = \{x_{i}\}_{i=1}^{n}$ parameterized by $\Theta_{n} = \{\theta_{i}\}_{i=1}^{n}$ drawn from $G$,  $n$ records can be classified into $k$ classes based on how they (the records) share parameters in $\Theta_{n}$. Then, by using the definition of Chinese Restaurant Process (CRP)~\cite{Teh:10}, class label $y_{n+1}$ for a new record $x_{n+1}$ can be predicted as follows:

\begin{eqnarray}
P(y_{n+1} = l_{j} | Y_{n}) \propto  \frac{n_{j}}{\alpha + n } \nonumber \\
P(y_{n+1} = l_{k + 1} | Y_{n}) \propto \frac{\alpha}{\alpha + n } 
\end{eqnarray}
where  $l_{j}$ is one of the labels for existing classes,   $j \in \{1, ..., k\}$,  and $l_{k+1}$ denotes the label of a new class. 
According to CRP, the probability of assigning a new incoming record to an existing class $l_{j}$ is proportional to the size of that class $n_{j}$, and the probability of generating an emerging class is proportional to the concentration parameter $\alpha$.

A DP Gaussian mixture model is obtained when each record $x_{i} \in {\rm I \!R}^{h}$ is generated from a Gaussian distribution whose parameter $\theta_{i} = \{\mu_{i}, \Sigma_{i}\}$ is drawn from $G$. Note that we assume our collected streaming records generated by INNMF step has the property of unimodality. Thus, we use a normally distributed data model, which can model unimodal class distributions fairly well. Next, we present the Dirichlet Process Gaussian Mixture Model (DPGMM) as below:

\begin{eqnarray}
x_{i} \sim N(x_{i} | \theta_{i}) \nonumber \\
\theta_{i} = \{\mu_{i}, \Sigma_{i}\} \sim G \nonumber \\
G \sim DP(\alpha, H)  
\end{eqnarray}

In DPGMM, each class component is modeled using a single Gaussian distribution. Due to the discreteness of the distribution $G$, records sharing the same parameter $\theta$ are considered as belonging to the same class. The base distribution $H$ in DPGMM is a conjugate Normal $\times$ Invert Wishart (NIW) prior, which is defined as follows:

\begin{eqnarray}
H &=& \text{NIW}(\mu_{0}, \Sigma_{0}, \kappa, m) \nonumber \\
&=& \mathcal{N}(\mu|\mu_{0}, \frac{\Sigma}{\kappa}) \times W^{-1}(\Sigma|\Sigma_{0}, m) 
\end{eqnarray}
where $\mu_{0}$ is the prior mean and $\kappa$ is a scaling constant that controls the deviation of the class conditional mean vectors from the prior mean. The parameter $\Sigma_{0}$ is a positive definite matrix that encodes our prior belief about the expected $\Sigma$. The parameter $m$ is a scalar that is negatively correlated with the degrees of freedom. For a given record $x_{i}$, by integrating out its corresponding parameters $\mu_{i}$ and $\Sigma_{i}$, its posterior predictive distribution for a Gaussian data model and NIW prior can be obtained in the form of multivariate student-t distribution~\cite{Anderson:84}:

\begin{eqnarray}
&p(x_{i} | y_{i} = l_{j}) = T(x_{i} | \overline{\mu}_{j}, \overline{\Sigma}_{j}, \overline{v}_{j})& \nonumber \\
\label{eq:stud-t}
&\overline{\mu}_{j} = \frac{n_{j}\mu_{j} + \kappa\mu_0}{n_{j}+\kappa}& \nonumber \\
&\scalebox{0.75}{$\overline{\Sigma}_{j} = \frac{n_{j}+\kappa+1}{(n_{j}+\kappa)(n_{j}+m+1-h)}\left(\Sigma_0 + (n_{j}-1)S_{j}+\frac{n_{j}\kappa}{n_{j}+\kappa} (\mu_{0} - \mu_{j}) (\mu_{0}- \mu_{j})^T\right)$}&  \nonumber \\ 
&\overline{v}_{j} = n_{j}+m+1-h &
\end{eqnarray}
where $\mu_{j}$ and $S_{j}$ are sample mean and sample covariance matrix for the class $l_{j}$. $\overline{\mu}_{j}$ is a $h \times 1$ mean vector, $\overline{\Sigma}_{j}$ is a $h \times h$ scale matrix, and $\overline{v}_{j}$ is the degree of freedom of the obtained
multivariate student-t distribution.

\subsection{Online Inference By One-Pass Gibbs Sampler}~\label{sec:gibbs}

Given $X_{n}$ (initially available records in vector representation using NNMF), $Y_{n}$ (known
labels of records in $X_n$), $\tilde{X}_{i-1}$ (the first $(i - 1)$ records observed online), and
$\tilde{Y}_{i-1}$ (the predicted labels of records in $\tilde{X}_{i-1}$), our goal is to evaluate the
conditional posterior probability of class indicator variable of $i$'th online record $\tilde{x}_{i}$ as soon as the record appears online. If
$\tilde{y}_{i}$ is the class indicator variable of $\tilde{x}_{i}$, the conditional posterior
probability of $\tilde{y}_{i}$ can be derived using one-pass Gibbs sampler as below: 

\begin{eqnarray}
\label{eq:posterior}
&& p(\tilde{y}_{i} = l_{j} | \tilde{Y}_{i-1}, \tilde{X}_{i}, Y_{n}, X_{n}) \nonumber \\
&\propto&
\begin{cases}
       \frac{n_{j}}{\alpha + n + i - 1}T(\tilde{x}_{i} | \overline{\mu}_{j}, \overline{\Sigma}_{j}, \overline{v}_{j})     \text{ if } 
     j \in \{1, ..., k + \tilde{k}_{i-1}\}  \\
      \frac{\alpha}{\alpha + n + i - 1}T(\tilde{x}_{i}),   \text{  if  }  
        j = k + \tilde{k}_{i-1} + 1 \\
\end{cases}      
\end{eqnarray} 

From Equation~\ref{eq:posterior}, the conditional posterior probability of $\tilde{y}_{i}$ depends on the posterior predictive likelihood of $\tilde{x}_{i}$ in the form of multivariate student-t distribution and CRP prior of the corresponding class component. Specifically, the incoming online record $\tilde{x}_{i}$ belongs to one of the existing classes with probability proportional to $\frac{n_{j}}{\alpha + n + i - 1}T(\tilde{x}_{i} | \overline{\mu}_{j}, \overline{\Sigma}_{j}, \overline{v}_{j})$, and a new class with probability proportional to $\frac{\alpha}{\alpha + n + i -1}T(\tilde{x}_{i})$. 
Note that $T(\tilde{x}_{i})$ is another multivariate student-t distribution by setting all sufficient statistics in Equation~\ref{eq:stud-t} to empty sets.

The pseudo-code of the proposed one-pass Gibbs sampler for online name disambiguation is summarized in Algorithm~\ref{alg:2}. Given a collection of $n$ records initially available in training set $X_{n}$, and their corresponding true label information $Y_{n}$, we aim to predict the class indicator variables of $n_{u}$ records sequentially observed online. Specifically, from line 2-6,  we utilize the conditional posterior probability shown in Equation~\ref{eq:posterior} to decide the class assignment of each online record, denoted as $\tilde{y}_{i}$. After processing all $n_{u}$ online records, we return the predicted class set $\hat{Y}_{pred}$ for final evaluation in line 7.

\begin{algorithm}
\renewcommand{\algorithmicrequire}{\textbf{Input:}}
\renewcommand{\algorithmicensure}{\textbf{Output:}}
\caption{One-Pass Gibbs Sampler for Online Name Disambiguation}
\label{alg:2} 
\begin{algorithmic}[1]
\REQUIRE $X_{n}$, $Y_{n}$, $n_{u}$
\ENSURE Final label prediction set $\hat{Y}_{pred}$
\STATE Initialize $\tilde{Y}_{0} = \emptyset$ 
\FOR {i = 1 to $n_{u}$}
   \STATE $\tilde{y}_{i} \sim p(\tilde{y}_{i} | \tilde{Y}_{i-1}, \tilde{X}_{i}, Y_{n}, X_{n})$
   \STATE $\tilde{Y}_{i} \gets \tilde{Y}_{i-1} \cup \{\tilde{y}_{i}\}$
   \STATE $\hat{Y}_{pred} \gets \hat{Y}_{pred} \cup \{\tilde{y}_{i}\}$
 \ENDFOR
\STATE \textbf{return} $\hat{Y}_{pred}$
\end{algorithmic}  
\end{algorithm}

\subsection{Online Inference By Particle Filtering}~\label{sec:SISR}

For the one-pass Gibbs sampler, the accumulated classification error from all mislabeled online records is propagated and eventually the model is likely to diverge from its true posterior distribution leading to poor online disambiguation performance.
To address this issue, we develop a Sequential Importance Sampling with Resampling (SISR)~\cite{Doucet.Godsill.ea:00} technique, also known as particle filtering in the literature. 
In contrast to a one-pass Gibbs sampler, particle filtering employs a set of particles, whose weights can be incrementally updated as new records appear online. Each particle maintains class configurations of all observed online records and the weight of a particle indicates how well the particle fits the data. Resampling ensures that particles with high weights are more likely to be replicated and the ones with low weights are more likely to be eliminated. By keeping a diverse set of class configurations, particle filtering allows for more effective exploration of the state-space and often generates a better local optimum than that would be obtained by a one-pass Gibbs sampler.

Specifically, in the particle filtering framework, for each online record, we approximate its true class posterior distribution by a discrete distribution defined by a set of particles and their weights, which can be incrementally updated without having access to all past records. Mathematically, we are interested in predicting $\tilde{Y}_{i}$, i.e., the class labels for all $\tilde{X}_{i}$ at the time $\tilde{x}_{i}$ appears online. The prediction can be done by finding the expectation of the posterior distribution of class indicator variables, namely $E_{p(\tilde{Y}_{i} | \tilde{Y}_{i-1}, \tilde{X}_{i}, Y_{n}, X_{n})}\Big[\tilde{Y}_{i}\Big]$. By using an importance function $q(\tilde{Y}_{i} | \tilde{Y}_{i-1}, \tilde{X}_{i}, Y_{n}, X_{n})$ to sample particles, in which case the $E_{p(\tilde{Y}_{i} | \tilde{Y}_{i-1}, \tilde{X}_{i}, Y_{n}, X_{n})}\Big[\tilde{Y}_{i}\Big]$ can be approximated as below: 

\small
\begin{eqnarray}
~& E_{p(\tilde{Y}_{i} | \tilde{Y}_{i-1}, \tilde{X}_{i}, Y_{n}, X_{n})}\Big[\tilde{Y}_{i}\Big] \\
=& \int \tilde{Y}_{i} p(\tilde{Y}_{i} | \tilde{Y}_{i-1}, \tilde{X}_{i}, Y_{n}, X_{n})d\tilde{Y}_{i} \nonumber \\
=& \int \tilde{Y}_{i} W_{i}\Big(\tilde{Y}_{i}\Big)q(\tilde{Y}_{i} | \tilde{Y}_{i-1}, \tilde{X}_{i}, Y_{n}, X_{n})d\tilde{Y}_{i} \nonumber \\
\approx& \sum_{m = 1}^{M} \tilde{Y}_{i}^{m} W_{i}\Big(\tilde{Y}_{i}^{m}\Big) \delta_{\tilde{Y}_{i}^{m}}
\label{eq:mean}
\end{eqnarray}
\normalsize
 
where $M$ is the number of particles, $\tilde{Y}_{i}^{m}$ represents the class configurations of first $i$ online records in $m$-th particle, and $W_{i}\Big(\tilde{Y}_{i}^{m}\Big) = \frac{p(\tilde{Y}_{i}^{m} | \tilde{Y}_{i-1}^{m}, \tilde{X}_{i}, Y_{n}, X_{n})}{q(\tilde{Y}_{i}^{m} | \tilde{Y}_{i-1}^{m}, \tilde{X}_{i}, Y_{n}, X_{n})}$ is the corresponding weight of the $m$-th particle when $i$-th online record is observed. Using the chain rule, the particle weights can be sequentially updated as follows:

\begin{eqnarray}
&W_{i} \Big(\tilde{Y}_{i}^{m}\Big) = \frac{p(\tilde{Y}_{i}^{m} | \tilde{Y}_{i-1}^{m}, \tilde{X}_{i}, Y_{n}, X_{n})}{q(\tilde{Y}_{i}^{m} | \tilde{Y}_{i-1}^{m}, \tilde{X}_{i}, Y_{n}, X_{n})}& \nonumber \\
&= W_{i-1}\Big(\tilde{Y}_{i-1}^{m}\Big) \frac{p(\tilde{x}_{i} | \tilde{Y}_{i}^{m}, \tilde{X}_{i-1}, Y_{n}, X_{n}) p(\tilde{y}_{i} | \tilde{Y}_{i-1}^{m}, Y_{n})}{p(\tilde{x}_{i} |\tilde{Y}_{i-1}^{m}, \tilde{X}_{i-1}, Y_{n}, X_{n}) q(\tilde{y}_{i} | \tilde{Y}_{i-1}^{m}, \tilde{X}_{i}, Y_{n}, X_{n})}&
\label{eq:weight}
\end{eqnarray}

To further simplify the formula, we set importance function to be CRP prior of the class indicator variable $\tilde{y}_{i}$, namely $q(\tilde{y}_{i} | \tilde{Y}_{i-1}^{m}, \tilde{X}_{i}, Y_{n}, X_{n}) = p(\tilde{y}_{i} | \tilde{Y}_{i-1}^{m}, Y_{n})$. Furthermore, $p(\tilde{x}_{i} |\tilde{Y}_{i-1}^{m}, \tilde{X}_{i-1}, Y_{n}, X_{n})$ is constant with respect to $\tilde{Y}_{i}^{m}$. Thus the weight update formula in Equation~\ref{eq:weight} can be simplified as below:

\begin{equation}
W_{i}\Big(\tilde{Y}_{i}^{m}\Big) \propto W_{i-1}\Big(\tilde{Y}_{i-1}^{m}\Big) p(\tilde{x}_{i} | \tilde{Y}_{i}^{m}, \tilde{X}_{i-1}, Y_{n}, X_{n})
\label{eq:weight-simple}
\end{equation}

In Equation~\ref{eq:weight-simple}, $p(\tilde{x}_{i} | \tilde{Y}_{i}^{m}, \tilde{X}_{i-1}, Y_{n}, X_{n})$ is a multivariate student-t distribution under the DPGMM. Thus at the time online record $\tilde{x}_{i}$ appears, the weights of all $M$ particles can be updated and then  normalized, namely $W_{i}\Big(\tilde{Y}_{i}^{m}\Big) = \frac{W_{i}(\tilde{Y}_{i}^{m})}{\sum_{p=1}^{M}W_{i}\Big(\tilde{Y}_{i}^{p}\Big)}$, into a discrete probability distribution to approximate the true conditional posterior distribution of its class indicator variable $\tilde{y}_{i}$.

In order to alleviate the problem of particle degeneracy and avoid the situation that all but only a few particle weights are close to zero, we add a stratified resampling step as suggested in \cite{Doucet.Godsill.ea:00}. The general philosophy of this resampling step is to replicate particles with high weights and eliminate particles with low weights.  At the time $\tilde{x}_{i}$ appears online, we use the weight update formula~\ref{eq:weight-simple} to compute weights of all particles. Then we calculate an estimate of the effective number of particles, which is defined as $ENP = \frac{1}{\sum_{m=1}^{M}[W_{i}(\tilde{Y}_{i}^{m})]^{2}}$. If this value is less than a given threshold, we perform resampling. Specifically, we draw $M$ particles from the current particle set (with replacement) with probabilities proportional to their weights and then we replace the current particle set with the new one. Meanwhile, we reset the weights of all particles to a uniform weight, which is $\frac{1}{M}$. In summary, in the particle filtering framework, after sampling particles using the CRP prior and updating particle weights as in Equation~\ref{eq:weight-simple}, we either retain weighted particles, in which case the weights are accumulated over time, or we resample particles so that they have uniform weights. 

\begin{algorithm}
\renewcommand{\algorithmicrequire}{\textbf{Input:}}
\renewcommand{\algorithmicensure}{\textbf{Output:}}
\caption{Particle Filtering Algorithm for Online Name Disambiguation}
\label{alg:1} 
\begin{algorithmic}[1]
\REQUIRE $X_{n}$, $Y_{n}$, $n_{u}$, $M$, $ENP_{thr}$
\ENSURE Final label prediction set $\hat{Y}_{pred}$
\STATE Initialize $\tilde{Y}_{0}^{m} = \emptyset$ and $W_{0}(\tilde{Y}_{0}^{m}) = \frac{1}{M}$ for all $m \in \{1, 2, ..., M\}$
\FOR {i = 1 to $n_{u}$}
  \FOR {m = 1 to $M$}
    \STATE $\tilde{y}_{i} \sim p(\tilde{y}_{i} |\tilde{Y}_{i-1}^{m}, Y_{n})$
    \STATE $\tilde{Y}_{i}^{m} \gets \tilde{Y}_{i-1}^{m} \cup \{\tilde{y}_{i}\}$
    \STATE $W_{i}\Big(\tilde{Y}_{i}^{m}\Big) \propto W_{i-1}\Big(\tilde{Y}_{i-1}^{m}\Big) p(\tilde{x}_{i} | \tilde{Y}_{i}^{m}, \tilde{X}_{i-1}, Y_{n}, X_{n})$
  \ENDFOR
  \STATE Normalize particle weights
  \IF {$ENP \le ENP_{thr}$}
    \FOR {m = 1 to $M$}
      \STATE Resample $\tilde{Y}_{i}^{m}$ with probability $\propto$ $W_{i}\Big(\tilde{Y}_{i}^{m}\Big)$
      \STATE $W_{i}\Big(\tilde{Y}_{i}^{m}\Big) = \frac{1}{M}$
    \ENDFOR
  \ENDIF
  \STATE Aggregate the particle weights based on all $M$ particle class labels and predict the label of $\tilde{x}_{i}$ with maximum weights, namely $\hat{Y}_{pred} \gets \hat{Y}_{pred} \cup \{\hat{y}_{i}\}$
\ENDFOR
\STATE \textbf{return} $\hat{Y}_{pred}$
\end{algorithmic}  
\end{algorithm}

The pseudo-code of the proposed particle filtering algorithm for online name disambiguation is summarized in Algorithm~\ref{alg:1}. 
Specifically, in line 1, we initialize the class configurations of all particles as $\emptyset$ and their weights as uniform weights. From line 2-8, we perform particle sampling and weight update steps. In line 9, if the effective number of particles criteria $ENP$ is below a threshold $ENP_{thr}$, we perform stratified resampling as shown in line 10-14. For the final prediction of class indicator variable for online record $\tilde{x}_{i}$, we sum the particle weights based on the particle class labels assigned to $\tilde{x}_{i}$  and choose the class label with the maximum weights as the final class label, denoted as $\hat{y}_{i}$ in line 15. Finally, we return the predicted class set $\hat{Y}_{pred}$ for evaluation. In contrast to particle filtering in Algorithm~\ref{alg:1}, one-pass Gibbs sampler shown in Algorithm~\ref{alg:2} can be approximately considered as a special but very restricted case of particle filtering with only one particle by setting the number of particles $M = 1$.

%
%
%
%

\section{Active Online Name Disambiguation}~\label{sec:active}

Based on our developed particle filtering based online inference algorithm, we propose an active learning framework for online name disambiguation, which mainly consists of the two steps as below:

\textbf{Active Selection:} The objective of this step is to identify records with most uncertain disambiguation results based on the posterior probability and seek user feedback for these cases for true label information. Specifically, for each online record, we first estimate its class conditional posterior probability using particle filtering (Section~\ref{sec:SISR}), then we use an entropy-based criteria to quantify the confidence of the tentative disambiguation result. Let the  probability of an online record belonging to class $l_{j}$ be $p_{j}$, and total number of predicted classes among all particle configurations be $|J|$. Then the entropy can be calculated as $-\sum_{j=1}^{|J|}p_{j}log p_{j}$. Note that the range of computed entropy values is between $0$ and $log|J|$, where $0$ means the disambiguation prediction is most confident and $log|J|$ means the disambiguation prediction is least confident. If the entropy is larger than a user-defined threshold, i.e., $\tau * log |J|$ ($0 \le \tau \le 1$), we consider the disambiguation result as uncertain and seek user feedback to obtain true class assignment for this particular record. 

\textbf{Model Update:} 
The goal of this step is to refine the conditional posterior probability of class indicator variable of current online record when its true label is offered by users. Specifically, we first use this ground-truth label information to update class configurations of all particles with respect to this particular online record and then refine the weight of each particle using Equation~\ref{eq:weight-simple}.

\section{Experimental Results}

In the experiment, we consider bibliographic data in a temporal stream format and disambiguate authors by partitioning their records (papers) into homogeneous groups. Specifically, we compare our proposed Bayesian non-exhaustive classification framework with various existing methods to demonstrate its superiority over those methods for performing online name disambiguation task. Furthermore, we also demonstrate the usages of proposed active online name disambiguation on real-world name reference.

\subsection{Datasets}

\begin{table}[t!]
\centering
\scalebox{0.95}{
\begin{tabular}{c c c c}
\toprule
Name  & \# Records & \# Attributes & \# Distinct  \\ 
Reference & & & Authors \\
\midrule
Kai Zhang & 66 & 488 & 24 \\
Bo Liu & 124 & 749 & 47 \\
Jing Zhang & 231 & 1456 & 85 \\
Yong Chen & 84 & 551 & 25 \\
Yu Zhang & 235 & 1440 & 72 \\
Hao Wang & 178 & 1074 & 48 \\
Wei Xu & 153 & 1037 & 48 \\
Lei Wang & 308 & 1819 & 112 \\
Bin Li & 181 & 1142 & 60 \\
Ning Zhang & 127 & 744 & 33 \\
Feng Liu & 149 & 919 & 32 \\
Lei Chen & 196 & 1052 & 40 \\
David Brown & 61 & 437 & 25 \\
Yang Wang & 195 & 1227 & 55 \\
Gang Chen & 178 & 1049 & 47 \\
X. Zhang & 62 & 601 & 40 \\
Yun Wang & 46 & 360 & 19 \\
Z. Wang & 47 & 498 & 38 \\
Bing Liu & 182 & 897 & 18 \\
Yang Yu & 71 & 444 & 19 \\
Ji Zhang & 64 & 398 & 16 \\
Bin Yu & 105 & 600 & 17 \\
Lu Liu & 58 & 425 & 17 \\
Ke Chen & 107 & 603 & 16 \\
Gang Luo & 47 & 270 & 9 \\
\bottomrule
\end{tabular}}
\caption{Arnetminer name disambiguation dataset}
\label{tab:dataset}
\vspace{-0.10in}
\end{table}

A key challenge for the evaluation of name disambiguation task is the lack of availability of
labeled datasets from diverse application domains. In recent years, the bibliographic repository site, Arnetminer~\footnote{\url{https://aminer.org/disambiguation}} has published several ambiguous author name 
references along with respective ground truths (paper list of each real-life person), which we use for 
evaluation. Specifically we use $25$ highly ambiguous (having
a larger number of distinct authors for a given name) name references and show the performance of
our method on these name references. The statistics of each name reference
are shown in Table~\ref{tab:dataset}. In this table,
we show the number of records, the number of binary attributes (explained in Section~\ref{sec:preprocessing}) and the number of distinct authors associated with that name
reference. It is important to understand that the online name disambiguation model is built on a name
reference, not on a source dataset, like Arnetminer as a whole, so each name reference is a distinct dataset on which the evaluation is performed.

\subsection{Competing Methods}

In order to illustrate the merit of our proposed approach, we compare our model with the following benchmark techniques. Among these the first two are existing state-of-the- art online name disambiguation methods, and the latter two are baselines that we have designed.

\begin{enumerate}

\item \textbf{Qian's Method~\cite{Qian.Zheng.ea:15}} 
Given the collection of training records initially available, for a new record, Qian's method computes class conditional probabilities for existing classes. This approach assumes that all the attributes are independent and the procedure of probability computation is based on the occurrence count of each attribute in all records of each class. 
Then the computed probability is compared with a pre-defined threshold value to determine whether the newly added record should be assigned to an existing class, or to a new class not yet included in the previous data. 

\item \textbf{Khabsa's Method~\cite{Khabsa.Giles.ea:15}}
Given the collection of training records initially available this approach first computes the $\epsilon$-neighborhood density for each online sequentially observed record. The $\epsilon$-neighborhood density of a new record is considered as the set of records within $\epsilon$ euclidean distance from that record. Then if the neighborhood is sparse, the new record is assigned to a new class. Otherwise, it is classified  into the existing class that contains the most records in the $\epsilon$-neighborhood of the new record.  

\item \textbf{BernouNaive-HAC:} In this baseline, we first model the data with a multivariate Bernoulli distribution (features are binary, so Bernoulli distribution is used) and train a Naive Bayes classifier. This classifier returns class conditional probabilities for each record in the test set which we use as meta features in a hierarchical agglomerative clustering (HAC) framework.

\item \textbf{NNMF-SVM-HAC:} We perform NNMF on our binary feature matrix and use the coefficients returned by NNMF to train a linear SVM. Class conditional probabilities for each test record are used as meta features in a hierarchical agglomerative clustering (HAC) framework the same way described above.
\end{enumerate}

\subsection{Experimental Setting and Implementation}

For each of the $25$ name references, we aim to build a separate model to classify the online records belonging to existing classes represented in the training set, as well as identifying records belonging to emerging classes not represented in the training set. In particular, we first train the model using the training set initially available,
then we add the records in the test set one-by-one in order to simulate new incoming streaming data. The  train and test
partition is based on the temporal order of each record in the dataset. To be more precise, we put the most recent $T_{0}$ years' records into the test set and the records from earlier years into the initially available training set. Furthermore, we verify how the performance of our proposed model varies as we tune the value of $T_{0}$. In the experiment, we set $T_{0}$ as $2$ and $3$. For the evaluation metric, we use mean-F1 measure~\cite{Zaki.Wagner:14},
which is unweighted average of F1-measure of individual classes. 
The range of mean-F1 measure is between $0$ and $1$, and a higher value indicates better disambiguation performance.

Our proposed Bayesian non-exhaustive classification framework has a few tunable parameters. Among them, the set of prior parameters $(\Sigma_{0}, \mu_{0}, m, \kappa)$ in the base distribution of NIW can be learned from the training set. For example, we use the mean of training set to estimate $\mu_{0}$, and set $\Sigma_{0}$ to be the pooled covariance matrix as suggested in~\cite{Greene.William:89}.
For $m$ and $\kappa$, we use vague priors for fixing their values to $h + 100$ and $100$ respectively. In addition to that, there are three additional user-defined parameters in our proposed framework. Specifically, we set latent dimension $h$ in INNMF as $10$, concentration parameter $\alpha$ in Dirichlet Process as $100$, and number of particles $M$ to be $100$. 
Finally, in particle filtering, when the effective number of particles is below $\frac{M}{2}$ as suggested by~\cite{Doucet.Godsill.ea:00},  we perform resampling.

For all the competing methods, we use identical set of features (before dimensionality reduction). 
We vary the probability 
threshold value of Qian's method and $\epsilon$ value of Khabsa's method by cross validation on the
training dataset.
and select the ones that obtain the best disambiguation performance in terms of Mean-F1 score.  
For BernouNaive-HAC and NNMF-SVM-HAC methods, during the hierarchical agglomerative clustering step, 
we tune the number of 
clusters in training set by cross validation in order to get the best disambiguation result. 

For both data processing and model implementation, we write our own code in Python and use
NLTK, NumPy, SciPy, scikit-learn, and filterpy libraries for data cleaning, linear algebra and machine learning operations. 
We run all the experiments on a 2.1 GHz Machine with $8$GB memory running Linux operating system. 

%
%
%
%
%
%
\begin{table*}[t!]
\centering
\scalebox{0.85}{
\begin{tabular}{c| c c c c| c c | c c c c | c}
\toprule
Name & \# train & \# test & \# emerge & \# emerge & {\bf Gibbs Sampler} & {\bf Particle Filter} & BernouNaive- & NNMF- & Qian's & Khabsa's & Improv. \\
Reference  & records          & records     & records    & classes  &   & & HAC & SVM-HAC& Method~\cite{Qian.Zheng.ea:15} & Method~\cite{Khabsa.Giles.ea:15}  & \\
\midrule
Kai Zhang  &  42  &  24  &  15  &  8  &  0.633 (0.041)  &  \textbf{0.661 (0.013)}  &  0.605  &  0.621  &  0.619  &  0.518  &  6.4\% \\
Bo Liu  &  99  &  25  &  11  &  8  &  0.716 (0.033)  &  \textbf{0.804 (0.011)}  &  0.733  &  0.719  &  0.714  &  0.559  &  9.7\% \\
Jing Zhang  &  121  &  110  &  56  &  35  &  0.591 (0.028)  &  \textbf{0.639 (0.008)}  &  0.554  &  0.566  &  0.590  &  0.631  &  1.3\% \\
Yong Chen  &  70  &  14  &  5  &  5  &  \textbf{0.889 (0.016)}  &  0.807 (0.006)  &  0.852  &  0.794  &  0.848  &  0.833  &  4.3\% \\
Yu Zhang  &  124  &  111  &  62  &  30  &  0.535 (0.013)  &  \textbf{0.678 (0.008)}  &  0.498  &  0.516  &  0.515  &  0.502  &  31.4\% \\
Hao Wang  &  148  &  30  &  9  &  8  &  \textbf{0.747 (0.026)}  &  0.672 (0.009)  &  0.635  &  0.639  &  0.702  &  0.581  &  6.4\% \\
Wei Xu  &  127  &  26  &  11  &  10  &  0.844 (0.033)  &  \textbf{0.892 (0.012)}  &  0.811  &  0.750  &  0.767  &  0.689   &  10.0\% \\
Lei Wang  &  245  &  63  &  28  &  24  &   0.705 (0.012)  &  \textbf{0.722 (0.007)}  &  0.701  &  0.708  &  0.703  &  0.620  &  2.0\% \\
Bin Li  &  154  &  27  &  11  &  9  &  0.807 (0.029)  &  \textbf{0.865 (0.011)}   &  0.775  &  0.733  &  0.775  &  0.743  &  11.6\% \\
Feng Liu  &  104  &  45  &  6  &  5  &  0.589 (0.031)  &  \textbf{0.719 (0.022)}  &  0.501  &  0.499  &  0.399  &  0.339  &  43.5\% \\
Lei Chen  &  96  &  100  &  24  &  18  &  0.356 (0.043)  &  0.438 (0.012)  &  \textbf{0.646}  &  0.527  &  0.430  &  0.222  &  -32.2\% \\
Ning Zhang  &  97  &  30  &  16  &  12  &  0.635 (0.021)  &  \textbf{0.713 (0.018)}  &  0.669  &  0.685  &  0.647  &  0.608  &  4.1\% \\
David Brown  &  48  &  13  &  4  &  3  &  0.839 (0.019)  &  \textbf{0.937 (0.006)}  &  0.904   &  0.593  &  0.816  &  0.450  &  3.7\% \\
Yang Wang  &  118  &  77  &  38  &  20  &  0.469 (0.033)  &  \textbf{0.698 (0.009)}  &  0.513  &  0.549  &  0.325  &  0.440  &  27.1\% \\
Gang Chen  &  113  &  65  &  20  &  14  &  \textbf{0.821 (0.004)}  &  0.816 (0.012)  &  0.474  &  0.467  &  0.451  &  0.357  &  73.2\% \\
X. Zhang  &  54  &  8  &  5  &  5  &  0.969 (0.018)  &  \textbf{1.0 (0.011)}  &  0.593  &  0.485  &  0.952  &  0.222  &   5.0\% \\
Yun Wang  &  31  &  15  &  6  &  6  &  0.680 (0.011)  &  \textbf{0.762 (0.005)}  &  0.512  &  0.479  &  0.644  &  0.358  &   18.3\% \\
Z. Wang  &  41  &  6  &  5  &  4  &  0.884 (0.023)  &  \textbf{0.906 (0.003)}  &  0.693  &   0.712  &  0.889   &  0.701  &  1.9\% \\
Bing Liu  &  156  &  26  &  4  &  4  &  0.495 (0.009)  &  \textbf{0.727 (0.004)}  & 0.318  &  0.466  &  0.356  &  0.406  &  56.0\%  \\
Yang Yu  &  51  &  20  &  6  &  6  &  0.503 (0.013)  &  0.648 (0.003)  &  0.499  &  0.523  &  \textbf{0.684}  &  0.493  &   -5.3\% \\
Ji Zhang  &  46  &  18  &  7  &  5  &  0.512 (0.024)  &  \textbf{0.616 (0.008)}  &  0.412  &  0.392  &  0.514  &  0.545  &  13.0\% \\
Bin Yu  &  87  &  18  &  7  &  4  &  0.469 (0.011)  &  \textbf{0.579 (0.004)}  &  0.488  &  0.526  &  0.564  &  0.540  &   2.7\% \\
Lu Liu  &  24  &  34  &  17  &  9  &  0.406 (0.012)  &  \textbf{0.497 (0.009)}  &  0.417  &  0.429  &  0.399  &  0.346  &  15.9\%  \\
Ke Chen  &  70  &  37  &  7  &  6  &  0.370 (0.012)  &  0.439 (0.005)  &  0.401  &  0.398  &  0.423  &  \textbf{0.501}  &  -12.4\% \\
Gang Luo  &  30  &  17  &  6  &  3  &  0.603 (0.022)  &  \textbf{0.865 (0.005)}  &  0.622  &  0.693  &  0.744  &  0.786  &   10.1\% \\
\bottomrule
\end{tabular}}
\caption{Comparison of mean-F1 values using records with most recent 2 years as test set. Paired t-test is conducted on all performance comparisons and it shows that all improvements are significant at the $0.05$ level.}
\label{tab:result1}
\vspace{-0.10in}
\end{table*}

\begin{table*}[t!]
\centering
\scalebox{0.85}{
\begin{tabular}{c| c c c c| c c | c c c c | c}
\toprule
Name & \# train & \# test & \# emerging & \# emerging & {\bf Gibbs Sampler} & {\bf Particle Filter} & BernouNaive- & NNMF- & Qian's & Khabsa's & Improv. \\
Reference  & records          & records     & records    & classes  &  & & HAC & SVM-HAC& Method~\cite{Qian.Zheng.ea:15} & Method~\cite{Khabsa.Giles.ea:15}  & \\
\midrule
Kai Zhang  &  27  &  39  &  20  &  10  &  0.602 (0.021)  &  \textbf{0.632 (0.011)}  &  0.503  &  0.584  &  0.520  &  0.510  &  8.2\% \\
Bo Liu  &  66  &  58  &  29  &   21  &  0.699 (0.011)  &  \textbf{0.767 (0.022)}  &  0.612  &  0.606  &  0.612  &  0.631  &  21.6\% \\
Jing Zhang  &  82  &  149  &  77  &  47  &  0.568 (0.022)  &  \textbf{0.601 (0.009)}  &  0.480  &  0.446  &  0.423  &  0.419  &  25.2\% \\
Yong Chen  &  54  &  30  &  12  &  8  &  0.775 (0.047)  &  \textbf{0.788 (0.021)}  &  0.615  &  0.701  &  0.615  &  0.545   &  12.4\% \\
Yu Zhang  &  87  &  148  &  71  &  38  &  0.457 (0.013)  &  \textbf{0.639 (0.019)}  &  0.445  &  0.615  &  0.447  &  0.412   &  3.9\% \\
Hao Wang  &  115  &  63  &  17  &  12  &  \textbf{0.698 (0.031)}  &  0.545 (0.011)  &  0.513  &  0.572  &  0.540  &  0.512  &  22.0\% \\
Wei Xu  &  101  &  52  &  17  &  14  &  0.734 (0.051)  &   \textbf{0.836 (0.028)}  &  0.683  &  0.603  &  0.635  &  0.586   & 22.4\% \\
Lei Wang  &  173  &  135  &  67  &  45  &  0.693 (0.044)  &  \textbf{0.701 (0.031)}  &  0.560  &  0.522  &  0.536  &  0.428  &  25.2\% \\
Bin Li  &  108  &  73  &  37  &  23  & 0.777 (0.009)  &  \textbf{0.828 (0.004)}  &  0.532  &  0.574  &  0.588  &  0.545  &  40.8\% \\
Feng Liu  &  70  &  79  &  9  &  8  &  0.545(0.017)  &  \textbf{0.618 (0.027)}  &  0.488   &  0.527  &  0.379  &  0.424  &  17.3\% \\
Lei Chen  &  65  &  131  &  39  &  25  &  0.332 (0.029)  &  0.382 (0.007)  &  \textbf{0.493}  &  0.447  &  0.398  &  0.176  &  -22.5\% \\
Ning Zhang  &  76  &  51  &  32  &  19  &  0.589 (0.034)  &  0.682 (0.019)  &  \textbf{0.744}  &  0.531  &  0.420  &  0.378  &  -8.3\% \\
David Brown  &  39  &  22  &  17  &  7  &  0.734 (0.008)  &  \textbf{0.899 (0.002)}  &  0.751  &  0.631  &  0.752  &  0.478  &  19.5\%  \\
Yang Wang  &  92  &  103  &  46  &  25  &  0.436 (0.012)  &  \textbf{0.627 (0.011)}  &  0.313  &  0.298  &  0.225  &  0.240  &  100.3\% \\
Gang Chen  &  89  &  89  &  27  &  19  &  \textbf{0.799 (0.008)}  &  0.737 (0.012)  &  0.347  &  0.407  &  0.383  &  0.221  &  96.3\% \\
X. Zhang  &  53  &  9  &  6  &  6  &  0.959 (0.016)  &  \textbf{0.992 (0.005)}  &  0.563  &  0.445  &  0.905  &  0.202  &   9.6\% \\
Yun Wang  &  25  &  21  &  17  &  9  &  0.535 (0.012)  &  \textbf{0.668 (0.006)}  &  0.501  &  0.438  &  0.567  &  0.385  &   17.8\% \\
Z. Wang  &  35  &  12  &  10  &  8  &  0.842 (0.013)  &  \textbf{0.894 (0.011)}  &  0.613  &   0.652  &  0.879   &  0.424  &  1.7\% \\
Bing Liu  &  141  &  41  &  9  &  5  &  0.415 (0.019)  &  \textbf{0.648 (0.006)}  & 0.307  &  0.481  &  0.286  &  0.371  &   34.7\% \\
Yang Yu  &  37  &  34  &  10  &  8  &  0.471 (0.014)  &  \textbf{0.539 (0.007)}  &  0.459  &  0.508  &  0.510  &  0.447  &   5.7\% \\
Ji Zhang  &  41  &  23  &  8  &  6  &  0.461 (0.017)  &  \textbf{0.591 (0.003)}  &  0.402  &  0.349  &  0.494  &  0.483  &  19.6\% \\
Bin Yu  &  80  &  25  &  11  &  47 &  0.430 (0.012)  &  \textbf{0.559 (0.008)}  &  0.461  &  0.539  &  0.423  &  0.463  &  3.7\% \\
Lu Liu  &  10  &  48  &  34  &  13  &  0.336 (0.013)  &  0.409 (0.011)  &  0.401  &  0.418  &  0.317  &  \textbf{0.426}  &   -4.0\% \\
Ke Chen  &  54  &  53  &  20  &  7  &  0.313 (0.012)  &  0.339 (0.009)  &  0.396  &  0.337  &  0.404  &  \textbf{0.442}  &  -23.3\% \\
Gang Luo  &  20  &  27  &  8  &  4  &  0.627 (0.021)  &  \textbf{0.810 (0.009)}  &  0.612  &  0.674  &  0.675  &  0.726  &   11.6\% \\
\bottomrule
\end{tabular}}
\caption{Comparison of mean-F1 values using records with most recent 3 years as test set. Paired t-test is conducted on all performance comparisons and it shows that all improvements are significant at the $0.05$ level}
\label{tab:result2}
\vspace{-0.10in}
\end{table*}

\subsection{Performance Comparison with Competing Methods}

Table~\ref{tab:result1} and Table~\ref{tab:result2} show the online name disambiguation performance between our proposed method and other competing methods for all $25$ name references. In both tables, \#train records and \#test records columns show the 
number of training and test records. \#emerge records column is the number of records in test set with their corresponding
classes not represented in the initial training set, and \#emerge classes column denotes the number of emerging classes not
represented in the training set. The columns 6-11 show the performance of a method using mean-F1 score for online disambiguation of records under a given name reference. The last column represents the overall improvement of our proposed method compared with the best competing method. Since both one-pass gibbs sampler and particle filtering based online inference techniques in our proposed online name disambiguation model are randomized algorithms, for each name reference we run the method $30$ times and report the average mean-F1 score. In addition, for our method, we also show the standard deviation in the parenthesis~\footnote{Standard deviation for other competing methods are not shown due to 
space limit.}. For better visual comparison, we highlight the best mean-F1 score of each name reference with bold-face font. 

%
%
If we compare the 25 datasets between the two tables, for higher $T_0$ value,  the number of training records decreases,
the number of test records, emerging records, and emerging classes increase. It makes
the online name disambiguation task in the first setting (2 years test split, i.e., $T_0=2$) easier than the
second setting ($T_0=3$) . This is reflected in the mean-F1
values of all the name references across both tables. For example, for the first name reference,
Kai Zhang, mean-F1 score of particle filtering across these two tables are 0.661 and 0.632 respectively. This performance
reduction is caused by the increasing number of emerging classes; 8 in Table~\ref{tab:result1}, and 10 in
Table~\ref{tab:result2}. Another reason is  decreasing number of training instances;
42 in Table~\ref{tab:result1}, and 27 in Table~\ref{tab:result2}.
As can be seen in both tables, our name disambiguation dataset contains a large number of emerging
records in the test data, all of these records will be misclassified with certainty by any traditional 
exhaustive name disambiguation methods. This is our main motivation for designing a non-exhaustive classification
framework for online name disambiguation task.

Now we compare our method with the four competing methods. As we observe, our proposed online name disambiguation model performs best for $22$ and $21$ name references (out of $25$) in Table~\ref{tab:result1} and Table~\ref{tab:result2}, respectively. Besides, the overall percentage improvement that our method delivers over the second best method is relatively large. For an example, consider the name reference ``Jing Zhang'' shown in Table~\ref{tab:result2}. This is a difficult online name disambiguation task as it contains a large number of emerge records in the test set ($77$ emerge records from $47$ emerge classes), thus any traditional classifier will misclassify all these emerge records with certainty. For our proposed particle filter and one-pass Gibbs sampler, it achieves $0.601$ and $0.568$ mean-F1 score for this name reference, respectively; whereas the best competing method for this name (BernouNaive-HAC) obtains only $0.480$, indicating a substantial improvement ($25.2\%$) by our method (particle filtering). The relatively good performance of the proposed method 
may be due to our non-exhaustive learning methodologies. It also suggests that the base distribution used by the proposed Dirichlet process prior model whose parameters are estimated using data from known classes can be generalized for the class distributions of unknown classes as well.  

When we compare between our proposed online inference approaches, particle filtering performs better than one-pass Gibbs sampler. The possible explanation could be that one-pass Gibbs sampler fails to maintain multiple local optimal solutions and prevent error propagations effectively during the online execution.  In comparison, particle filtering offers more accurate approximation of class posterior distribution for each online record, which leads to better mean-F1 performance across most of name references.

\begin{table}[t!]
\centering
\caption{Results of number of distinct real-life persons under our proposed Bayesian non-exhaustive classification framework using most recent 3 years' records as test set}
\scalebox{0.80}{
\begin{tabular}{c | c | c c}
\toprule
Name  & \# Actual Authors & \# Predicted Authors  & \# Predicted Authors \\
Reference & & (Gibbs Sampler) &  (Particle Filter) \\
\midrule
Kai Zhang & 15 & 19.1 $\pm$ 3.2 & 20.3 $\pm$ 3.7 \\
Bo Liu &  30 & 34.2 $\pm$ 2.0 & 31.3 $\pm$ 2.9 \\
Jing Zhang & 62 & 51.4 $\pm$ 4.8 & 56.3 $\pm$ 5.6 \\
Yong Chen & 13 & 14.2 $\pm$ 1.9 & 16.9 $\pm$ 2.2 \\
Yu Zhang & 55 & 60.3 $\pm$ 4.8 & 56.8 $\pm$ 4.3 \\
Hao Wang & 27 & 35.6 $\pm$ 2.3 & 33.8 $\pm$ 3.2 \\
Wei Xu & 27 & 30.2 $\pm$ 1.8 & 29.0 $\pm$ 2.5 \\
Lei Wang & 65 & 70.1 $\pm$ 4.6 & 68.5 $\pm$ 3.4 \\
Bin Li & 32 & 35.6 $\pm$ 4.1 & 37.0 $\pm$ 3.5 \\
Feng Liu & 26 & 29.7 $\pm$ 3.3 & 23.8 $\pm$ 2.9 \\
Lei Chen & 12 & 18.2 $\pm$ 2.9 & 14.5 $\pm$ 4.1 \\
Ning Zhang & 23 & 21.1 $\pm$ 1.7 & 24.8 $\pm$ 2.2 \\
David Brown & 11 & 13.4 $\pm$ 3.6 & 14.2 $\pm$ 1.8 \\
Yang Wang & 36 & 28.9 $\pm$ 4.5 & 35.2 $\pm$ 2.6 \\
Gang Chen & 28 & 30.2 $\pm$ 1.8 & 31.8 $\pm$ 2.0 \\
X. Zhang & 9 & 10.3 $\pm$ 3.7 & 12.1 $\pm$ 2.8 \\
Yun Wang & 12 & 14.3 $\pm$ 3.9 & 11.7 $\pm$ 0.9 \\
Z. Wang & 10 & 12.4 $\pm$ 3.7 & 11.6 $\pm$ 1.2 \\
Bing Liu & 8 & 10.6 $\pm$ 1.9 & 9.2 $\pm$ 2.2 \\
Yang Yu & 14 & 17.4 $\pm$ 5.7 & 15.8 $\pm$ 2.6 \\
Ji Zhang & 11 & 10.3 $\pm$ 6.9 & 12.1 $\pm$ 2.3\\
Bin Yu & 11 & 9.3 $\pm$ 2.8 & 12.3 $\pm$ 1.6 \\
Lu Liu & 15 & 12.1 $\pm$ 3.3 & 16.3 $\pm$ 1.5 \\
Ke Chen & 12 & 14.1 $\pm$ 2.4 & 13.2 $\pm$ 1.9 \\
Gang Luo & 5 & 7.9 $\pm$ 2.0 & 10.2 $\pm$ 2.7 \\
\bottomrule
\end{tabular}}
\label{tab:result3}
\vspace{-0.20in}
\end{table}

In contrast, among all the competing methods, Qian's method and Khabsa's method perform the worst 
as they fail to incorporate prior information about class distribution into the models and the results 
are very sensitive to the selections of threshold parameters. On the other hand both BernouNaive-HAC 
and NNMF-SVM-HAC operate in an off-line framework. Although for some name references mean-F1 scores obtained 
by these techniques are higher than our proposed method,  there is a clear trend favoring our proposed 
method over these methods---latter cannot explicitly identify streaming records of new ambiguous classes 
in an online setting.

Table~\ref{tab:result3} presents the result of automatic estimation of  number of distinct real-life persons in test set using both one-pass Gibbs sampler and particle filtering. From the table, as we observe, for most of name references, the predicted number of distinct persons are slightly larger than the actual ones. 
The possible explanation could be that our name disambiguation datasets contain skewed classes, and DPGMM tends to produce multiple components for each class. Despite that, as a remark, our predicted results are very close to the actual ones, which demonstrate the effectiveness of our proposed online name disambiguation framework
for estimating the number of actual real-life persons accurately under the non-exhaustive setup. 

\begin{figure}
\includegraphics[width = \linewidth]{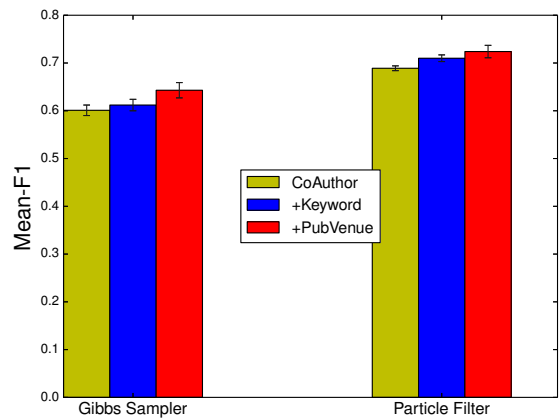}
\caption{Feature contribution analysis using most recent 2 years' publication records as test set. The results are averaged out over all $25$ name references for better visualization.}
\label{fig:feature-analysis}
\vspace{-0.1in}
\end{figure}

\subsection{Feature Contribution Analysis}

We investigate the contribution of each of the defined features (coauthor, keyword, venue) for the task of online name disambiguation.
Specifically, we first rank the individual features by their performance in terms of mean-F1 score, then add the features one by one in the order of their disambiguation power. In particular, we first use author-list, followed by keywords, and publication venue. 
In each step, we evaluate the performance of our proposed online name disambiguation method using the
most recent two years' publication records as test set. Figure~\ref{fig:feature-analysis} shows the mean-F1 value of our method with different feature combinations. As we can see from this figure, for both one-pass Gibbs sampler and particle filtering based online inference techniques, after adding each feature group we observe improvements in terms of mean-F1 score,  in which the results are averaged out over all the $25$ name references for better visualization.

\subsection{Study of Running Time}

A very desirable feature of our proposed Bayesian non-exhaustive classification model is its running time. For example, using the most recent two years' records as test set, on the name reference ``Kai Zhang" containing $66$ papers with $10$ latent dimensionality, it takes around $0.29$ and $1.09$ seconds on average to assign the test papers to different real-life authors for one-pass Gibbs sampler and particle filtering, respectively. For the name reference ``Lei Wang" with $308$ papers using same number of latent dimensionality, it takes around $1.95$ and $5.91$ seconds on average under the same setting. This suggests only a linear increase in computational time with respect to the number of records. However
in addition to number of records, the computational time depends on other factors, such as the latent dimensionality, the number of particles, and the number of classes generated, which in turn depends on the values of the hyperparameters used in the data model and concentration parameter in dirichlet process.

\begin{figure*}
\centering
\subfigure[Mean-F1 vs. latent dimension]{\label{fig:parameter1}\includegraphics[width=55mm]{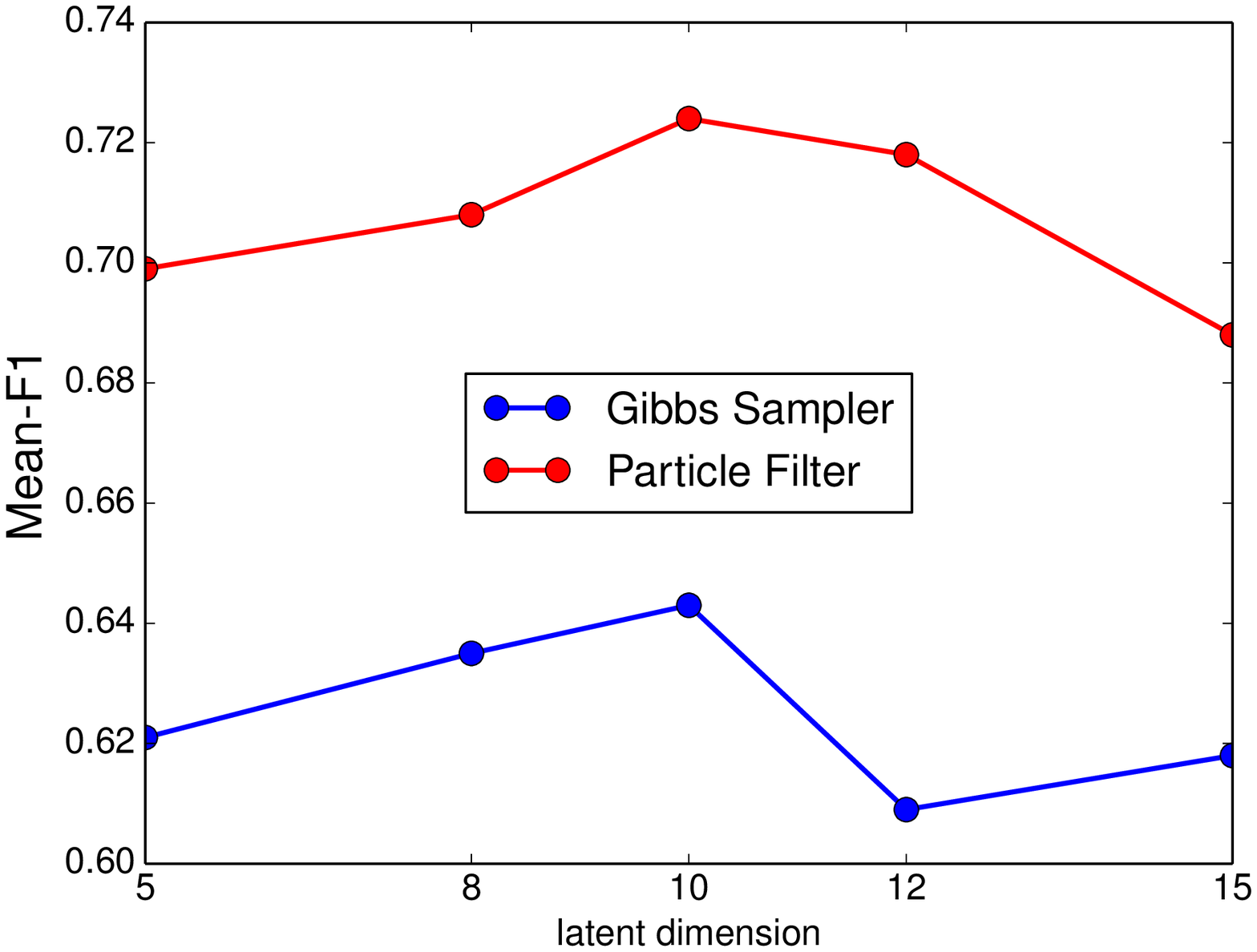}}
\subfigure[Mean-F1 vs. concentration parameter]{\label{fig:parameter2}\includegraphics[width=55mm]{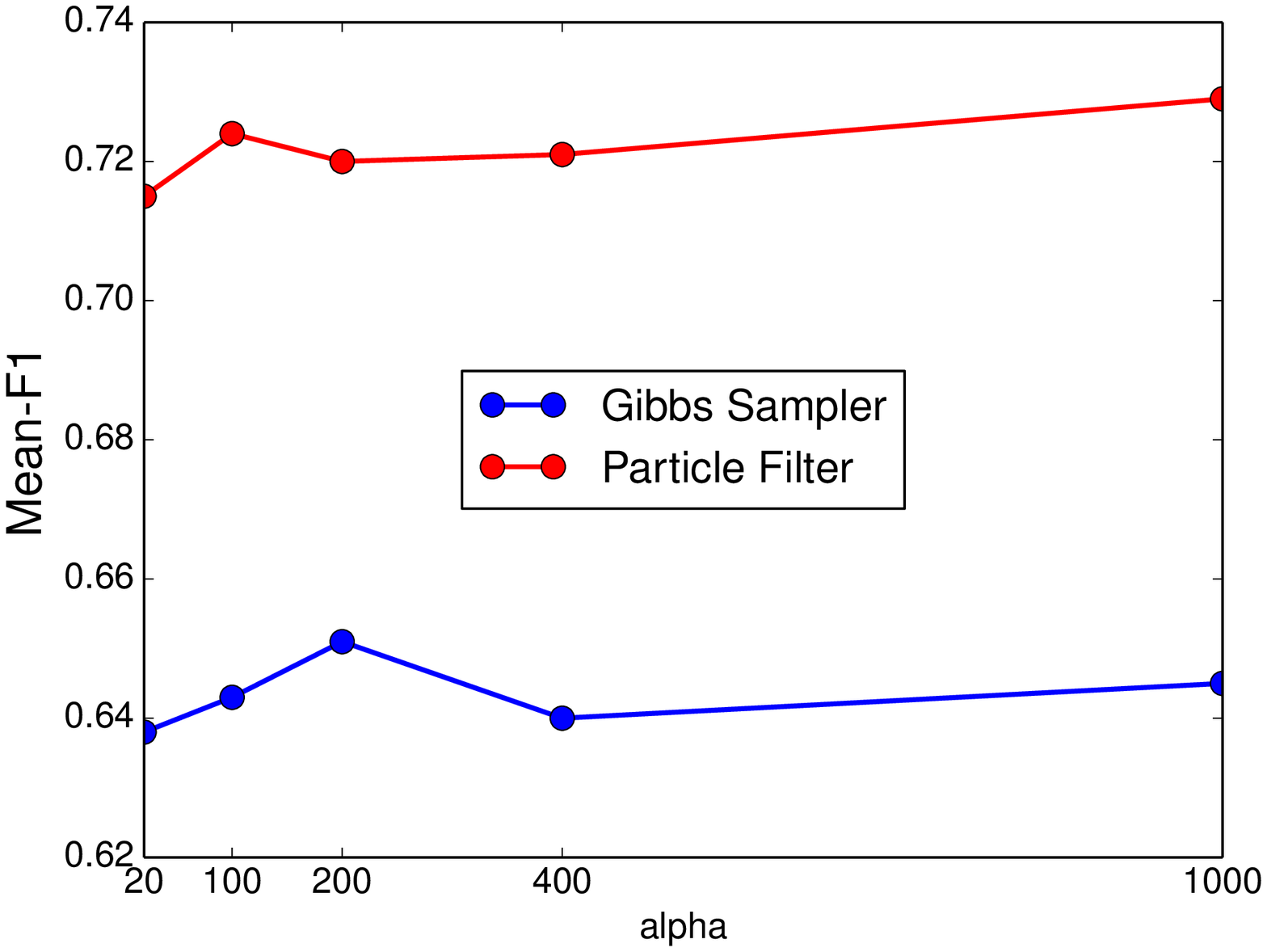}}
\subfigure[Mean-F1 vs. \# particles]{\label{fig:parameter3}\includegraphics[width=55mm]{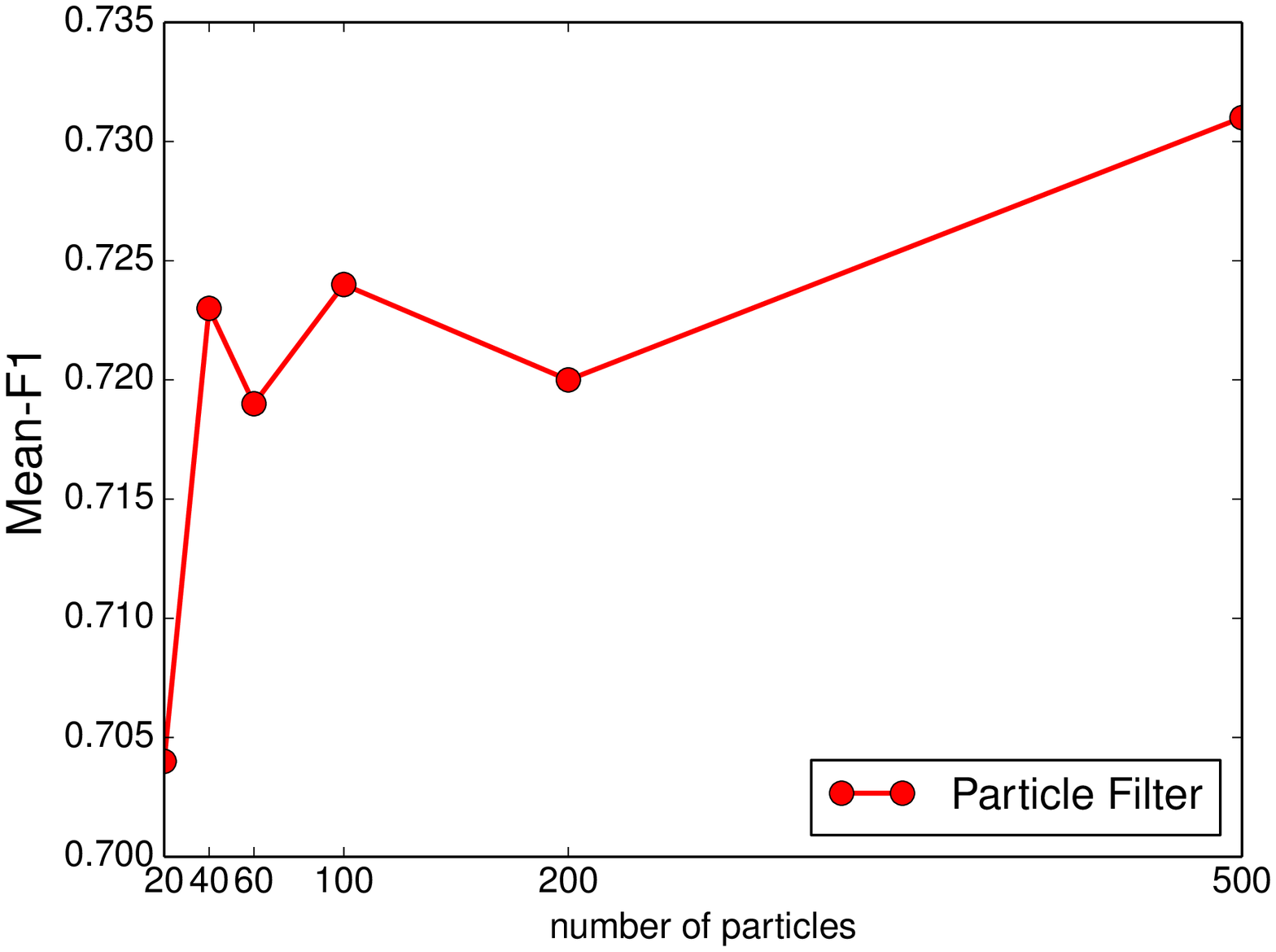}}
\caption{The effects of latent dimension $h$, concentration parameter $\alpha$ in Dirichlet process, and number of particles $M$ in particle filtering on the online name disambiguation performance using most recent 2 years' records as test set. The results are averaged out over all $25$ name references.}\label{fig:ps}
\vspace{-0.1in}
\end{figure*}

\subsection{Study of Parameter Sensitivity}~\label{sec:1}

In our proposed Bayesian non-exhaustive classification framework, there are three user-defined parameters, namely latent dimension $h$, concentration parameter $\alpha$ in Dirichlet process, and number of particles $M$ in particle filtering. In this section, we investigate the classifier performance with respect to these three parameter variations. The results are shown in Figures~\ref{fig:parameter1},~\ref{fig:parameter2},~\ref{fig:parameter3}. Specifically, first for latent dimension sensitivity, as we see from Figure~\ref{fig:parameter1}, for both online inference approaches, as the latent dimension increases, the online name disambiguation performance in terms of Mean-F1 first
increases and then decreases. The possible explanation is that
when the latent dimension is too small, the representation capability of the
latent feature is not sufficient and we may lose information. However, when
the latent dimension is too large, the proposed INNMF technique (details in Section~\ref{sec:preprocessing}) is too
complex and we may over-fit to the data. The second parameter $\alpha$ in the Dirichlet process prior model (Section~\ref{sec:DPGMM})
controls the probability of assigning an incoming record to a new class and it plays a critical role in the number 
of generated classes in the online name disambiguation process. As we can observe from Figure~\ref{fig:parameter2}, the classifier performance is robust 
with respect to different $\alpha$ values. Finally, Figure~\ref{fig:parameter3} shows that only a few number of particles is sufficient to have desirable classifier prediction (details in Section~\ref{sec:SISR}). 

\begin{figure}
\includegraphics[width = \linewidth]{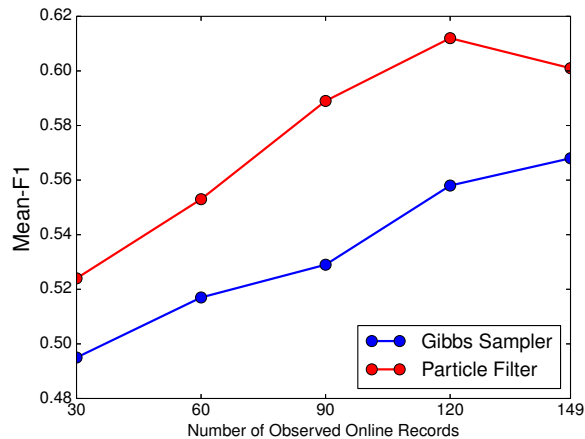}
\caption{Online name disambiguation performance over the number of observed online records on name reference ``Jing Zhang'' using most recent 3 years' records as test set.}
\label{fig:per_time}
\vspace{-0.1in}
\end{figure}

\subsection{Performance over the number of observed online records}~\label{sec:2}

We investigate the online name disambiguation performance over the number of sequentially observed records. We use the name reference
``Jing Zhang"~\footnote{We choose name reference ``Jing Zhang'' as a case study due to the fact that it contains largest number of test records ($149$) among all name references used in the experiment.~\label{refnote}} with its corresponding most recent 3 years' records as test set as a case study. Specifically, we evaluate the mean-F1 score as
we process $\{30, 60, 90, 120, 149\}$ test records. As we see from Figure~\ref{fig:per_time}, as more records are observed online, the overall disambiguation performance improves in both one-pass Gibbs sampler and particle filtering based online inference techniques. The results demonstrate that our proposed
learning model has self-adjusting capacity that accurately classify incoming online records to both novel and existing classes and effectively prevent error propagation during the online execution stage.

\begin{figure}
\centering
\includegraphics[width = \linewidth]{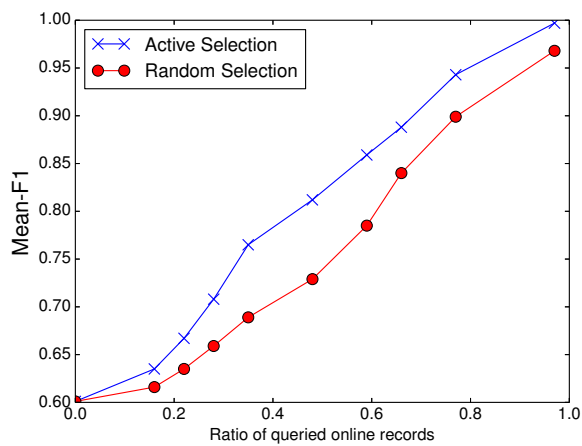}
\caption{Mean-F1 comparison between our proposed active selection and random selection with respect to different ratios of
queried online records on name reference ``Jing Zhang" using most recent 3 years' records as test set. The higher the curve, the better the performance.}
\label{fig:ac-learn}
\vspace{-0.2in}
\end{figure}

\subsection{Results of Active Online Name Disambiguation}~\label{sec:3}

Now we compare the results of our proposed particle filtering algorithm for online name disambiguation with active selection and random selection (randomly selecting a number of sequentially observed records to query the users for ground-truth). Specifically, for the user-defined interactiveness threshold  parameter $\tau$ (defined in Section~\ref{sec:active}), we set it to $\{0.1, 0.2, ..., 0.9, 1,0 \}$, and run our proposed active online name disambiguation method $20$ times under each $\tau$. Then we compute the average ratio of queried online records for different values of $\tau$. Note that the larger the value of $\tau$, the
fewer the number of queried online records. For the random selection, it queries an online record with probability $0 < p < 1$. In other words, for each online record, the method draws a value from a uniform distribution $U(0, 1)$. If the value is smaller than $p$, it queries the label. Otherwise, it does not. For a fair comparison, we set $p$ as the ratio of queried online records in our proposed interactive framework. The result on name reference ``Jing Zhang" is shown in Figure~\ref{fig:ac-learn}.

As we observe, for both active and random selection frameworks, compared to the no feedback scenario where $\tau$ is set to be $1.0$ and we don't query any online records, incorporating user feedback helps to improve online name disambiguation performance in terms of mean-F1 score. However, our proposed active selection framework is better than random selection consistently under different ratios of queried online records for performing active online name disambiguation. In particular, our proposed active online name disambiguation framework actively queries those records whose label information are uncertain.  
In contrast, the labels acquired by the random selection may be redundant and lead to the waste of labeling effort. Similar results are obtained for other name references as well.

\section{Conclusion and Future Work}~\label{sec:conclude}

To conclude, in this paper we present a Bayesian non-exhaustive classification framework for the task of online name disambiguation. Given sequentially observed online records, our proposed method classifies the incoming records into existing classes, as well as emerging classes 
by learning posterior probability of a Dirichlet process Gaussian mixture model. 
Our experimental results on bibliographic datasets demonstrate that the proposed method significantly outperforms the existing state-of-the-arts. As a real-life application, we propose an active online name disambiguation method to improve the prediction accuracy by exploiting user feedback. 

There are still rooms to improve the method proposed in this work. The data model used in this study is limited with
the Gaussian distribution. The proposed approach can be
extended to problems involving more flexible class distributions
by choosing a mixture model for each class
data and a hierarchical Dirichlet Process Prior model over class distributions. Another future work would be to use 
time-dependent Dirichlet process to incorporate temporal information into the prior model. 
Specifically,  for bibliographic data, a clear temporal trend exists; most people in academia start with $1-2$ papers 
per year,  and then increase this rate significantly during their career's peaks which diminish as 
they get closer to retirement. Incorporating such intuition into the model may also improve
the online name disambiguation performance substantially. 

\bibliographystyle{ACM-Reference-Format}
\balance
\bibliography{disambiguation}  

%
%
\end{document}